\documentclass{jpp}
\usepackage{graphicx}

\usepackage[utf8]{inputenc}
\usepackage[T1]{fontenc}
\usepackage{amsmath}

\usepackage{csquotes}

\shorttitle{Anisotropic Heating in Electron-only Magnetic Reconnection}
\shortauthor{J. Ren, G. Arr{\`o}, M. E. Innocenti and G. Lapenta}

\title{Anisotropic Heating and Parallel Heat Flux in Electron-only Magnetic Reconnection with Intense Guide Fields}

\author{Jincai Ren\aff{1}
  \corresp{\email{jincai.ren@kuleuven.be}},
  Giuseppe Arr{\`o}\aff{2},
  Maria Elena Innocenti\aff{3}
 \and Giovanni Lapenta\aff{1}}

\affiliation{\aff{1}Department of Mathematics, KU Leuven, Celestijnenlaan 200B, 3001 Leuven, Belgium
\aff{2}Los Alamos National Laboratory, Theoretical Division, Los Alamos, NM 87545, United States
\aff{3}Institut für Theoretische Physik, Ruhr-Universität Bochum, 44801 Bochum, Germany}

\begin{document}

\maketitle

\begin{abstract}
Electron-only reconnection is a process recently observed in the Earth's magnetosheath, where magnetic reconnection occurs at electron kinetic scales with only electrons involved. Electron-only reconnection is likely to have a significant impact on the energy conversion and dissipation of turbulence cascades at kinetic scales.
This paper investigates electron-only reconnection under different intensities of strong guide fields via two-dimensional fully kinetic Particle-in-Cell (PIC) simulations, focusing on electron heating. 
The simulations are initialized with a force-free current sheet equilibrium under various intensities of strong guide fields.
The quadrupolar structures of the out-of-plane magnetic field and asymmetric structures of density are found to become narrower in space as the guide field increases.
Electron velocities are considerably larger than ion velocities, and the motion of both species is notably affected by the strength of guide fields.
Similarly to previous experiments studies, electron temperature anisotropy along separatrices is observed, which is found to be mainly caused by the variations of parallel temperature. Both regions of anisotropy and parallel temperature increase/decrease along separatrices become thinner with increasing guide fields. Besides, we find a transition from a quadrupolar to a six-polar to an eight-polar structure in temperature anisotropy and parallel temperature as the guide field intensifies.
Non-Maxwellian electron velocity distribution functions (EVDFs) at different locations in the three simulations are observed.
Our results show that parallel electron velocity varies notably with different guide field intensities and strong parallel electron heat flux is observed.

\end{abstract}

\section{Introduction}

Magnetic reconnection is a ubiquitous phenomenon observed in space, fusion and laboratory plasmas. During this process, the magnetic field topology is modified and magnetic energy is converted into plasma kinetic and thermal energy \citep{biskamp2000magnetic,birn2007reconnection,hesse2020magnetic,lu2023energy}. Throughout the past decades, the standard ion-scale magnetic reconnection has been studied extensively through laboratory experiments, spacecraft observations, theoretical analysis and numerical simulations, and significant progress has been made within this domain \citep{innocenti2015evidence,hesse2020magnetic,ji2023laboratory}. The term \enquote{standard ion-coupled magnetic reconnection} here refers to a reconnection regime in which both ions and electrons are involved. Such reconnection processes take place in current sheets of size typically larger than ion scales and imply the transfer of energy from magnetic fields to both ions and electrons. As a result, both ions and electrons are accelerated out of the reconnection region, with each species displaying rapid outflow jets moving in opposite directions \citep{zweibel2009magnetic}.

On the other hand, a novel regime of magnetic reconnection, dubbed electron-only reconnection (e-rec hereafter), has been observed at scales smaller than ion scales, and only electrons are involved in this process. Thanks to the advancement of spacecraft in-situ diagnostics with unprecedented high resolution, e-rec has been observed in the Earth's magnetosheath by \citet{phan2018electron} for the first time. 
Given the tiny spatial and temporal scales characterizing e-rec, ions are unable to respond to the small-scale magnetic field dynamics and thus ion outflow jets existing in standard ion-coupled reconnection are absent in e-rec \citep{sharma2019transition,califano2020electron,arro2020statistical,shi2022laboratory,shi2022electron,shi2023using,lu2022electron,guan2023reconnection,roy2024energy}. Differently from its ion-scale counterpart, e-rec does not exhibit oppositely directed ion jets out of the reconnection region, and only electron outflows are observed. 
Furthermore, the thickness of the associated current sheet is of the order of the electron inertial length $d_e$ and the electron outflow speed is super-Alfvénic \citep{phan2018electron}.

Electron-scale current sheets connected with e-rec have been observed in satellite measurements and numerical simulations of turbulence and large-scale reconnection \citep{phan2018electron,stawarz2019properties,stawarz2022turbulence,arro2020statistical}, and they are likely to have a significant impact on the energy conversion and dissipation within turbulence environments like magnetosheath of the Earth \citep{califano2020electron,arro2020statistical,franci2022anisotropic}. It has been demonstrated that e-rec also develops as a secondary effect in the outflows of large-scale three-dimensional reconnection \citep{lapenta2015secondary,lapenta2022formation}, a phenomenon which has been confirmed by spacecraft observations \citep{zhong2021three}. 
\citet{sharma2019transition} studied the transition from ion-coupled reconnection to e-rec under conditions of plasma beta greater than one and strong guide fields (\(B_g=8B_{x0}\), hereafter $B_g$ and $B_{x0}$ represent the strength of the guide field and in-plane asymptotic reconnecting field, respectively). 
\citet{califano2020electron} and \citet{arro2020statistical} investigated the formation and roles of e-rec in plasma turbulence and analyzed the statistical properties of turbulent fluctuations in both the e-rec and the standard ion-coupled reconnection regimes. However, the particle heating in e-rec still remains largely unexplored.

Remarkable progress has been made in understanding two-dimensional anti-parallel reconnection during the past decades. Here, anti-parallel reconnection refers to a configuration in which two regions with oppositely directed magnetic fields are present in a plane and the out-of-plane component of magnetic fields (i.e., the guide field) is zero. In this case, reconnection happens at the neutral sheet between the oppositely directed magnetic fields.
It has been shown that the addition of the out-of-plane guide field alters the reconnection process dramatically, even when the guide field is small in magnitude. The presence of guide fields decreases reconnection rate, modifies the structure of diffusion region, and affects particle motion \citep{birn2007reconnection,ricci2004collisionless,hesse2006dissipation,stanier2015fast,stanier2015fluid,munoz2016non,wilson2016particle}. 
A density asymmetry is typically observed along separatrices in guide field reconnection \citep{kleva1995fast,pritchett2005onset,birn2007reconnection,lapenta2011bipolar,markidis2012three}. Usually, the density is enhanced on one separatrix but is reduced along the other. Besides, the reconnection current sheet typically tilts \citep{birn2007reconnection}. Consistently with the quadrupolar field configuration predicted by the Hall-reconnection model, a quadrupolar out-of-plane magnetic field perturbation generated by the in-plane electron current is observed in guide field reconnection \citep{pritchett2005onset}. Furthermore, tripolar magnetic field structures have also been reported in \citep{newman2022tripolar}. 
The guide field is also found to have a remarkable impact on the location of energy transfer and particle dynamics, including increasing the electron energy gain and affecting the electron heating in the diffusion region \citep{yi2020energy,guo2015localized}.
Up to now, the majority of guide field reconnection research has been focusing on the ion-coupled regime, typically involving weak or moderate strength guide fields \citep{horiuchi2014influence,mccubbin2022characterizing} (Here we regard the guide field with strength of \(B_g/B_{x0}<5\) as weak or moderate and the guide field with strength of \(B_g/B_{x0}\geq5\) as intense/strong).
E-rec in the presence of intense guide fields is still not well explored numerically, with remarkable exceptions \citep{sharma2019transition,guan2023reconnection, roy2024energy}.

The energy transfer between the electromagnetic fields and plasma influences both the kinetic and thermal energy of electrons during magnetic reconnection \citep{yang2017energy2,yang2017energy}. \citet{pucci2018energy} studied parallel and perpendicular heating in guide field reconnection with $B_g/B_{x0}$ ranging from 0 to 3 through 2.5D Particle-in-Cell (PIC) simulations.
More recently, \citet{shi2023using} has shown evidence of temperature anisotropy developing along separatrices in e-rec experiments with strong guide fields ($B_g/B_{x0}=25$), conducted in the context of the PHASMA (PHAse Space MApping facility) experiment \citep{shi2021alfvenic}. Their investigation suggests that the observed anisotropy arises from parallel electric field heating rather than by Fermi or betatron acceleration.
The anisotropy along the separatrices appearing in the PHASMA experiment exhibits a different pattern with respect to the anisotropy observed in previous small-scale reconnection simulations \citep{pucci2018energy}, an effect probably caused by the non-negligible Coulomb collisions taking place in the PHASMA experiment \citep{shi2023using}. 
However, the role of strong guide fields in influencing temperature anisotropy in e-rec has been poorly investigated from a numerical point of view.

Non-Maxwellian electron velocity distribution functions (EVDFs) have been observed in both simulations \citep{ng2011kinetic,munoz2014instabilities,munoz2016non,shuster2015spatiotemporal,shay2016kinetic} and satellite observations \citep{burch2016electron,lapenta2017origin,zhou2019observations} of standard ion-coupled magnetic reconnection. In \citet{egedal2008evidence} and \citet{egedal2013review}, the authors developed a trapping model to explain the acceleration and the temperature anisotropy typically observed in guide field reconnection, which has been validated by both spacecraft observations and numerical simulations. 
On the contrary, the kinetic features of e-rec are still poorly explored and the only experimental evidence for non-Maxwellian EVDFs comes from the PHASMA experiments \citep{shi2022laboratory,shi2022electron,shi2023using}.

In this work, we analyse three numerical simulations of e-rec with varying intense guide field strengths ($B_g/B_{x0}=5, 10$ and $20$, respectively), focusing on electron heating. The plasma \(\beta\) for electrons in these three cases are 0.48, 0.12 and 0.03, respectively.
The quadrupolar structures of the out-of-plane magnetic field perturbations are shown. Similarly to standard ion-coupled reconnection, we find asymmetric density structures for electrons and ions in e-rec. Electron outflow velocity is super-Alfvénic and ion outflow jets are absent, which is strong evidence that our simulations are in the electron-only regime. We also find that the increase of guide fields leads to the thinning of both the quadrupolar structure of the out-of-plane magnetic field and the asymmetric structure of density.
Electron temperature anisotropy along separatrices is observed in our e-rec simulations. Perpendicular temperature increase is found along one separatrix, while perpendicular temperature decreases along the other. 
Parallel temperature increase/decrease, on the other hand, is much stronger than perpendicular temperature variations and plays a major role in the anisotropy. Parallel temperature increase is on one separatrix, while parallel temperature decrease distributes along the other. Both regions of the anisotropy and parallel temperature increase/decrease along separatrices become narrower as the guide field intensifies. Moreover, the strength of guide fields has notable effects on the parallel temperature in the outflow regions. We find a transition from a quadrupolar to a six-polar to an eight-polar structure in temperature anisotropy and parallel temperature as the guide field increases.
We further analyse the non-Maxwellian EVDFs at different locations, especially regions with strong anisotropy. Our results show that the EVDFs are gyrotropic, especially in the last two cases with stronger guide fields. Moreover, parallel electron velocity varies notably with different guide field intensities and strong parallel electron heat flux is observed.

\section{Simulation setup}\label{part2}

The simulations are carried out in a two-dimensional geometry using the fully kinetic semi-implicit Particle-in-Cell code ECsim \citep{lapenta2017exactly,gonzalez2018performance}, which has been extensively used to simulate magnetic reconnection \citep{lapenta2017multiple}, plasma turbulence \citep{arro2022spectral}, fusion tokamak dynamics \citep{ren2024recent} and in solar wind instabilities \citep{micera2020particle}.
Given the high guide field regime, implying a low plasma \(\beta\), we initialize our simulations with a force-free current sheet equilibrium, which is unstable to the plasmoid instability.
In our setup, $x$ is the asymptotic direction of the in-plane magnetic field, $\mathrm{y}$ is the direction across the sheets and $z$ is the out-of-plane direction. 
We use a reduced (but still quite high) ion-to-electron mass ratio of $m_i/m_e=100$, so that the ratio between ion and electron inertial length is ${d_i}/{d_e}=10$. The simulation domain is a square box whose size is $L_x\times L_{\mathrm{y}} = 20d_{e}\times20d_{e} = 2d_i \times 2d_i$.
The $x$ component of the initial magnetic field profile is given by:
\begin{equation}
	B_x(\mathrm{y})=B_{x0}\left[tanh\left(\frac{\mathrm{y}-\mathrm{y}_1}{\delta_1}\right)-tanh\left(\frac{\mathrm{y}-\mathrm{y}_2}{\delta_2}\right)-1\right]
\end{equation}
Here $\mathrm{y}_1$ and $\mathrm{y}_2$ represent the location of the two current sheets, with $\mathrm{y}_1=L_{\mathrm{y}}/4$, $\mathrm{y}_2=3L_{\mathrm{y}}/4$. $\delta_1$ and $\delta_2$ are the thickness of two current sheets at $\mathrm{y}=\mathrm{y}_1$ and $\mathrm{y}=\mathrm{y}_2$, respectively. Here $B_{x0}=0.002$.
The thicknesses of the two current sheets are $\delta_1=0.1d_i=1d_e$ and $\delta_2=0.05d_i=0.5d_e$ ($d_i$ and $d_e$ are ion inertial length and electron inertial length, respectively) so both sheets are at electron kinetic scales. We find the evolution of reconnection in these two sheets have similar patterns except that the thinner (upper) sheet evolves faster, as expected \citep{zelenyi1979relativistic}.  We choose the upper sheet for the analysis and its evolution remains unaffected by the lower sheet, as the latter exhibits a slower development.
The out-of-plane magnetic field $B_z$ is given by:
\begin{equation}
	B_z(\mathrm{y})=\left[B_{g}^2+B_{x0}^2-B_x^2(\mathrm{y})\right]^{1/2}
\end{equation}
with $B_{g}$ denoting the asymptotic guide field, far from the current sheet. This magnetic field configuration implies a force-free equilibrium in two-dimensional geometry.
All three simulations are conducted with the same parameters, except for the asymptotic guide field strength $B_{g}$. It is chosen to be $B_{g}=0.01=5B_{x0}$ for simulation \textbf{A}, $B_{g}=0.02=10B_{x0}$ for simulation \textbf{B}, and $B_{g}=0.04=20B_{x0}$ for simulation \textbf{C}. These strengths are typically observed in the Earth's magnetosheath.
The initial current is calculated from Ampere's law (without displacement current) and all currents are carried by electrons. The thermal velocities of electrons and ions are $v_{th,e}=0.05\,c$ and $v_{th,i}= 0.002\,c$, respectively ($c$ is the speed of light). We do not add perturbations to our simulations, and reconnection, in such thin current sheets, starts as a plasmoid instability triggered by the numerical noise \citep{newman2022tripolar,pritchett2005onset}.

The simulation box is discretized with a mesh of $512\times 512$ cells. The grid step is about $\triangle x=\triangle \mathrm{y} \simeq 0.04d_{e}$. Both the electron gyroradius and electron inertial length are resolved with this spatial resolution for all cases. The time step is $dt=0.01\omega_{ce}^{-1},\,0.02\omega_{ce}^{-1},\,0.04\omega_{ce}^{-1}$ ($\omega_{ce}$ is electron cyclotron frequency) in the three simulations, respectively, resolving the electron cyclotron motion in all simulations. The number of particles per cell is $8192$ for each species.
Periodic boundary conditions are considered in all directions. 
Particles are initialized from Maxwellian distributions. The density is uniform for both species and quasi-neutrality is initially satisfied.
The ratio between the electron cyclotron frequency \(\omega_{ce}\) and the ion cyclotron frequency \(\omega_{ci}\) is $\omega_{ce}/\omega_{ci}=100$.

\section{Results}\label{part3}

\subsection{Reconnected flux and reconnection rate}\label{part3.1}

\begin{figure}
    \centering
    \includegraphics[scale=0.065]{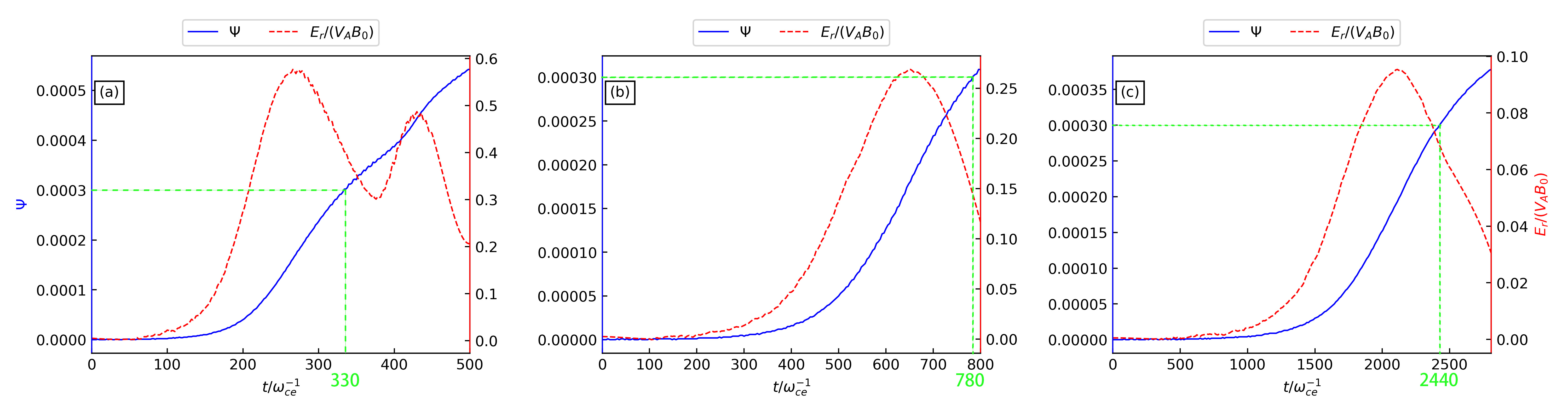}
    \caption{Reconnected flux and reconnection rate in the three cases: (a): case \textbf{A} (\(B_g=5B_{x0}\)); (b): case \textbf{B} (\(B_g=10B_{x0}\)); (c): case \textbf{C} (\(B_g=20B_{x0}\)). The green dashed lines represent the time snapshots used to compare the three simulations. These three snapshots have the same reconnected flux and roughly the same island size. The reconnection rate \(E_r\) is normalised by the inflow Alfvén speed \(V_A\) and in-plane asymptotic magnetic field \(B_0\). }
    \label{RateFlux}
\end{figure}

The reconnected flux and reconnection rate of simulation \textbf{A} are shown in Fig.~\ref{RateFlux} (a). The reconnected flux $\Psi$ is computed as the difference between the maximum and minimum of the out-of-plane component of vector potential $A_z$ along the upper current sheet. 
The reconnection rate \(E_r\) is calculated as $E_r=\partial \Psi/\partial t$ and is normalised by the in-plane asymptotic magnetic field \(B_0\) and the inflow Alfvén speed \(V_A=B_0/\sqrt{\mu_0 \rho}\). We find that reconnection goes through a slow \enquote{build-up} phase at first and this lasts until about $t=80\omega_{ce}^{-1}$. Then the reconnection rate increases steadily and reaches the first peak at $t=265\omega_{ce}^{-1}$. Afterwards, it decreases first and later returns to increasing from $t=380\omega_{ce}^{-1}$, which is caused by the coalescence of magnetic islands. A second peak is then reached, followed by a new decrease. It is noted that the first peak rate is larger.

The reconnected flux and reconnection rate of simulations \textbf{B} and \textbf{C} are illustrated in Fig.~\ref{RateFlux} (b) and (c), respectively. They have similar patterns as case \textbf{A}. The reconnection is very slow initially, and then after the \enquote{build-up} phase, both the reconnected flux and the reconnection rate increase gradually. Then the rate reaches a peak value and decreases later while the reconnected flux still increases. However, case \textbf{B} and \textbf{C} do not exhibit a second peak, unlike case \textbf{A}. This is because simulations B and C are not run to the point where magnetic islands coalesce. The coalescence does not affect our results since we carry out the analysis at an earlier time (as shown below). We find that the reconnection is delayed with large guide fields. The reconnection rate starts to increase from $t\approx 100\omega_{ce}^{-1}$ in case \textbf{A}, $t\approx 300\omega_{ce}^{-1}$ in case \textbf{B} and $t\approx 1000\omega_{ce}^{-1}$ in case \textbf{C}, respectively. 
Furthermore, it is noted that the maximum reconnection rate decreases when the guide field increases.
These results prove that large guide fields delay the onset of reconnection and decrease the reconnection rate, which is analogous to previous ion-coupled reconnection works \citep{hesse2002structure,ricci2004collisionless,munoz2016non}. 
In the following sections, we will analyze and compare the three simulations when the plasmoid instability is well-developed. The times chosen for the comparison are indicated by the green dashed lines in Fig.~\ref{RateFlux}, and we select those times because they correspond to the same reconnected flux. Since the reconnected flux is a measure of the islands' size, the islands at these three times have roughly the same size.

\subsection{Structure of the reconnection region}\label{part3.2}

\begin{figure}
    \centering
    \includegraphics[scale=0.395]{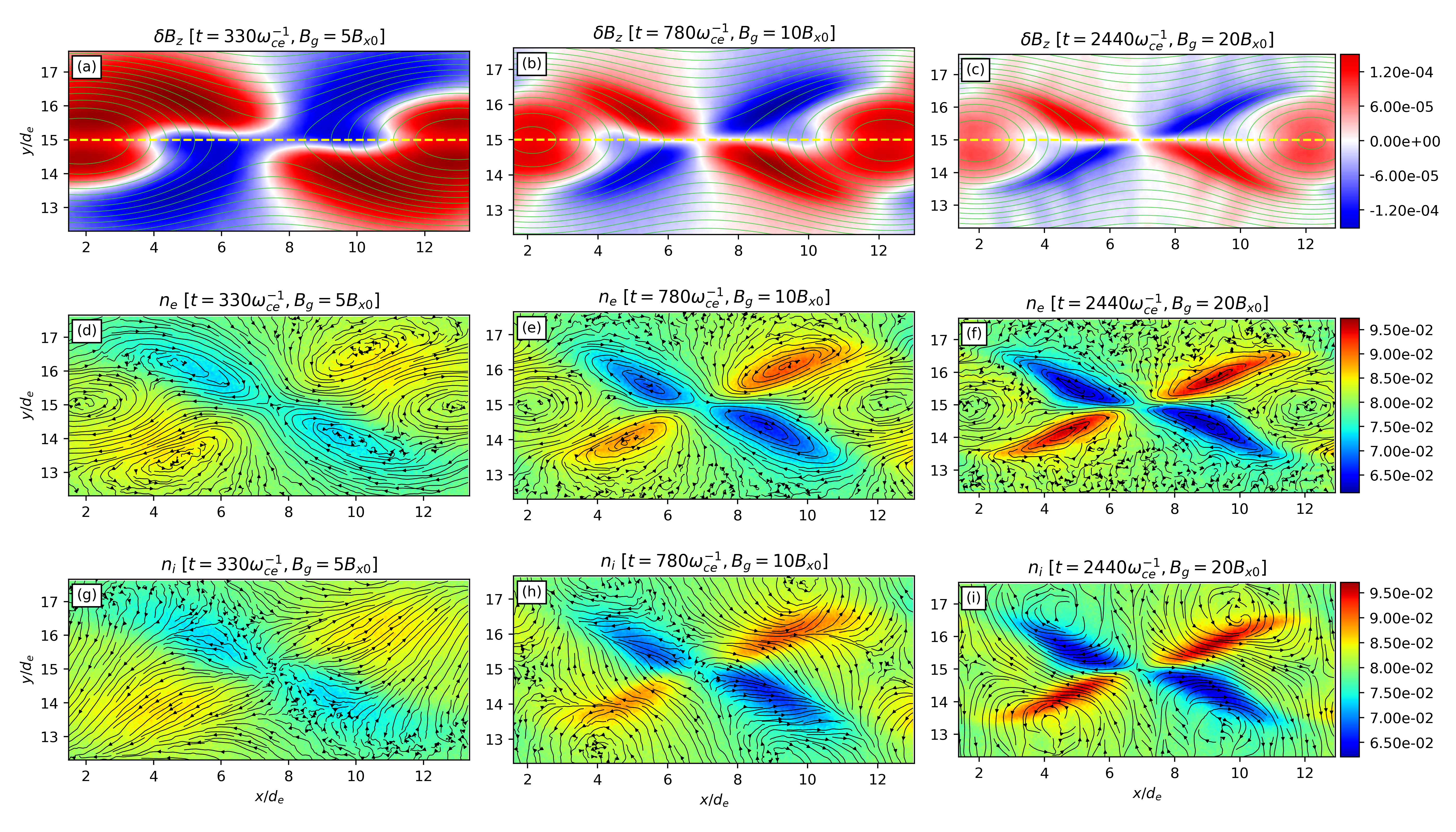}\\
    \vspace{-8.5pt}
    \includegraphics[scale=0.39]{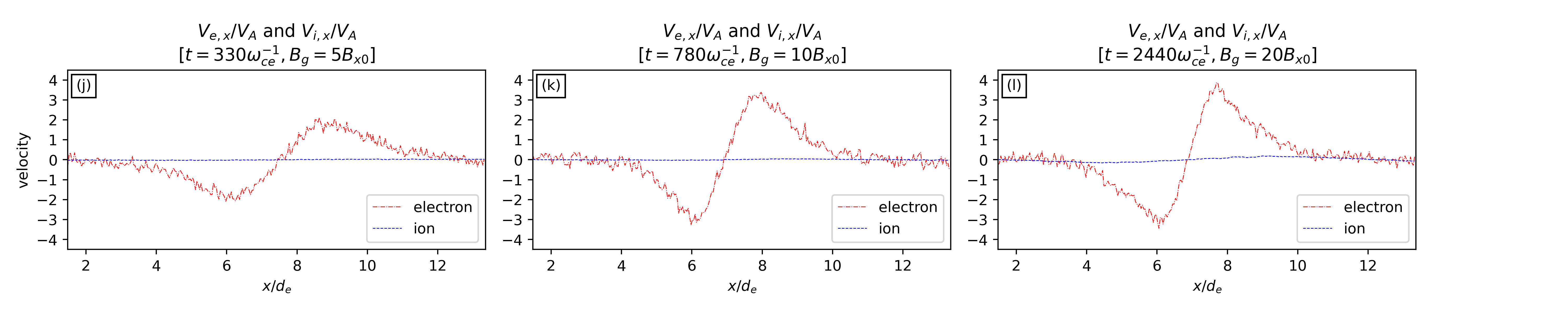}
    \caption{First column: \(t=330 \omega_{ce}^{-1}\) in case \textbf{A} (\(B_g=5B_{x0}\)); second column: \(t=780 \omega_{ce}^{-1}\) in case \textbf{B} (\(B_g=10B_{x0}\)); third column: \(t=2440 \omega_{ce}^{-1}\) in case \textbf{C} (\(B_g=20B_{x0}\)).
    First row: Structure of the out-of-plane magnetic field perturbations \(\delta B_z\); second row: electron density \(n_e\) and electron velocity streamlines; third row: ion density \(n_i\) and ion velocity streamlines; fourth row: electron velocity \(V_{e,x}\) and ion velocity \(V_{i,x}\) along x at \(\mathrm{y}=15.0\). In each row of first three rows, the three figures share the same colorbar, which is located on the far right of the row.}
    \label{dBnV}
\end{figure}

In Fig.~\ref{dBnV} (a) to (c), we compare the out-of-plane magnetic field fluctuations $\delta B_z=B_z(t)-B_z(t=0)$ of the three runs at the times indicated in Fig.~\ref{RateFlux}.
Here, $B_z(t)$ denotes the out-of-plane magnetic field at time $t$ (so $B_z(t=0)$ represents the initial out-of-plane magnetic field). Solid lines represent the in-plane component of magnetic field lines and we notice the presence of a well-developed reconnection region in all three runs, with magnetic islands of comparable size.
We find that $B_z$ increases along the separatrix in the upper left and lower right quadrants (where \(\delta B_z>0\)), whereas it decreases along the other separatrix in the upper right and lower left quadrants (where \(\delta B_z<0\)).
This result is the same as the quadrupolar structure in the Hall reconnection model \citep{zweibel2009magnetic,munoz2015fully} and is similar to \citet{sharma2019transition}. It is noted that the perturbation is relatively small and would not change the sign of the total out-of-plane magnetic field.
In addition, we also find the structure of $\delta B_z$ varies as the guide field increases. In case \textbf{A}, $\delta B_z$ is less localized, exhibiting wide spatial variations, as shown in Fig.~\ref{dBnV} (a). Furthermore, $\delta B_z$ is negative at the X-point and its neighboring areas along the neutral sheet. However, in case \textbf{B} and \textbf{C}, the structure of \(\delta B_z\) along the separatrices becomes more localized and appears to be nearly anti-symmetric with respect to the neutral sheet, as shown in Fig.~\ref{dBnV} (b) and (c), respectively. 
This localization effect of the quadrupolar structure is sensitive to the guide fields, as it becomes more and more prominent with the increase of the guide field strength, as illustrated in Fig.\ref{dBnV} (a) to (c). 

Asymmetric density structures have been observed around the separatrices in standard ion-scale guide field reconnection \citep{kleva1995fast,pritchett2004three,pritchett2005onset,birn2007reconnection,lapenta2011bipolar,markidis2012three}. 
It is pointed out that electron parallel motion accelerated by the parallel electric field (along newly reconnected magnetic field lines) contributes to density cavities (low density regions) along one separatrix and density enhancements (high density regions) along the other separatrix. Similar density structures are also obtained in our simulations. Fig.~\ref{dBnV} (d) to (f) illustrate the electron density (represented by the colored contour plots) in the three simulations. 
Electron density enhancements develop along the separatrix in the lower left and upper right quadrants. In contrast, electron density cavities reside on the other separatrix in the upper left and lower right quadrants.
Analogously, ion density exhibits similar asymmetric structures, as shown in Fig.~\ref{dBnV} (g) to (i). This is because ions move to neutralize electrons, albeit at significantly smaller velocities. The electron and ion density structures roughly match, which implies that the net charge is small.
The different in-plane motion of electrons and ions in all cases is compared in Fig.~\ref{dBnV} (d) to (i), where streamlines with arrows illustrate the in-plane component of the fluid velocities of the two species. The electron fluid motion exhibits patterns correlated to the out-of-plane magnetic field fluctuations and density asymmetry structures as expected, since these structures are due to the still magnetized electrons carrying the magnetic field in the out-of-plane direction. On the other hand, the in-plane ion motion is significantly different as ions decouple from the reconnection process. The motion separation between electrons and ions produces in-plane net currents, inducing the out-of-plane magnetic field perturbations observed in Fig.~\ref{dBnV} (a) to (c), in accordance with Ampere's law. It is noted that the spatial extent of both quadrupolar structures and density enhancements/cavities is confined to a few electron inertial lengths in e-rec, typically less than 1 ion inertial length (\(d_i\)). This contrasts with the standard ion-coupled reconnection regime, where such structures generally span several ion inertial lengths \citep{pritchett2004three}.
As expected, electron and ion velocities also differ in magnitude. Fig.~\ref{dBnV} (j) to (l) illustrate the velocities of electrons and ions (normalised by the Alfvén speed \(V_A\)) in the x-direction along \(\mathrm{y}=15.0\) (marked by the yellow dashed lines in Fig.~\ref{dBnV} (a) to (c)) in all three cases. The electron outflow jets are observed in the results, while the ion outflow jets existing in the standard ion-coupled reconnection are absent (the ion velocities are still nearly 0 quite far away from the X-point along the current sheet), proving that reconnection is in the electron-only regime. The results are consistent with \citet{sharma2019transition}. Furthermore, this also implies that the in-plane currents are predominantly carried by electrons.

The guide field has noticeable effects on density structures and particle motion patterns. With increasing guide field strength, the regions of density enhancement and cavity become narrower, as illustrated in Fig.~\ref{dBnV} (d) to (f) for electrons and (g) to (i) for ions, respectively.
Additionally, the maximum (minimum) values in the density enhancement (cavity) regions increase (decrease) with growing guide fields. 
The narrowing effect also appears in the electron flow results. The in-plane electron velocity exhibits a rotational pattern along local density cavities and enhancement regions. It turns more localized with increasing guide fields, as depicted in Fig.~\ref{dBnV} (d) to (f). 
As for the ions, we find that in the lowest guide field case, their in-plane motion shows variations at larger scales with respect to electrons, but as the guide field increases, we see some vortical ion motion around the reconnection site, albeit not much correlated to density variations. We also observe that the electron velocity increases slightly with larger \(B_g\).
We find the density variations along the enhancement/cavity regions are relatively modest compared to the background density, a feature notably different from the dramatic variations observed in previous standard ion-coupled reconnection studies \citep{pritchett2004three,pritchett2005onset}.
Furthermore, we observe that the density does not show substantial variations inside the islands in the outflow. This is a remarkable difference with respect to ion-coupled reconnection, where density typically increases inside magnetic islands, as reported in \citep{pritchett2004three,pritchett2005onset}.

\subsection{Anisotropic electron heating}\label{part3.3}

\begin{figure}
    \centering
    \includegraphics[scale=0.395]{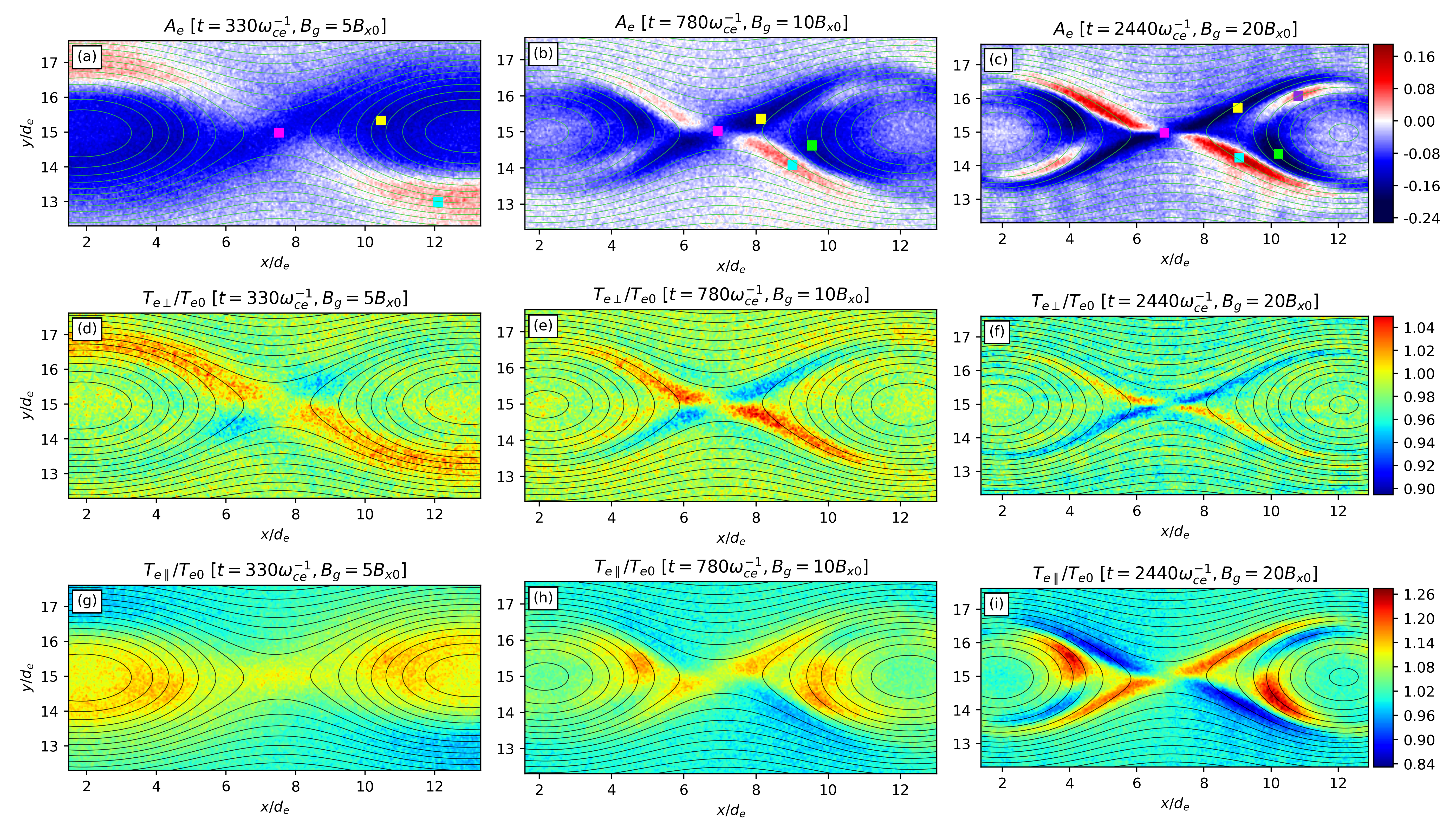}
    \caption{Electron temperature anisotropy \(A_e\), normalized perpendicular and parallel temperatures: \(T_{e, \perp}/T_{e,0}\) and \(T_{e, \parallel}/T_{e,0}\) in the three cases. \(T_{e,0}\) is the initial electron temperature. The colored boxes represent the positions at which EVDFs are calculated in these three cases. In each row, the three figures share the same colorbar, which is located on the far right of the row.}
    \label{AT}
\end{figure}

Previous studies on guide field reconnection have revealed anisotropic heating observed at different locations in both experiments and simulations \citep{guo2017energy,pucci2018energy,shi2022laboratory,shi2022electron,shi2023using}. 
In this study, we investigate this problem with a broader range of intense guide fields.
Fig.~\ref{AT} (a) to (c) depict the electron temperature anisotropy $A_e=T_{e,\perp}/T_{e,\parallel}-1$ at the same time as Fig.~\ref{dBnV} for the three simulations, with $T_{e,\perp}$ and $T_{e,\parallel}$ representing the electron temperature perpendicular and parallel to the local magnetic field, respectively. 
Positive values of $A_e$ indicate that perpendicular temperature exceeds parallel temperature locally (caused by perpendicular heating and/or parallel cooling), whereas negative values of $A_e$ imply that parallel temperature is larger than perpendicular temperature locally (caused by parallel heating and/or perpendicular cooling). Our analysis reveals a consistent presence of significant anisotropy along separatrices in all cases. Specifically, negative anisotropy (depicted by blue regions) is observed along the separatrix in the lower left and upper right quadrants and at the X-point, which means that parallel temperature is dominant in these regions.
Conversely, positive anisotropy (illustrated by red regions) is situated along the separatrix in the upper left and lower right quadrants, indicating relatively larger perpendicular temperature in these areas.
However, notable differences are observed with varying guide field strengths. As the guide field increases, the highest value of anisotropy increases, the anisotropic regions narrow, and the zones of positive anisotropy shift closer to the X-point (although \(A_e\) remains negative at the X-point). 
Apart from the separatrices, pronounced anisotropy also emerges in the outflow regions. In case \textbf{A} (Fig.~\ref{AT} (a)), negative anisotropy dominates within the outflow regions. Nevertheless, negative anisotropy reduces at a certain location (the faint white region near \(x\approx 11, \mathrm{y}\approx 16\) shown in Fig.~\ref{AT} (b)) in the outflow regions when the guide field increases. In case \textbf{C}, the anisotropy at this location is observed to become positive, as illustrated in Fig.~\ref{AT} (c). 

To gain insight into the underlying mechanisms driving the observed temperature anisotropy, we examine the parallel and perpendicular electron temperature, 
which are defined as $T_{e,\parallel} = \textbf{b} \cdot \textbf{P}_e \cdot \textbf{b} / n_e$ and $T_{e,\perp} = (3\, T_e - T_{e,\parallel})/2$, respectively. Here $\textbf{P}_e$ is the electron pressure tensor, $\textbf{b}=\textbf{B}/|\textbf{B}|$ is the magnetic field unit vector, and $T_e=Tr(\textbf{P}_e)/n_e$ is the isotropic electron temperature (with $Tr(\cdot)$ being the trace operator).
In Fig.~\ref{AT} (d) to (f), we show the electron perpendicular temperature ratio \(T_{e, \perp}/T_{e, 0}\) (\(T_{e, 0}\) represents the initial isotropic electron temperature) in all three cases. 
We find perpendicular temperature increase (\(T_{e, \perp}/T_{e, 0} >1\)) along the separatrix in the upper left and lower right quadrants.
In contrast, perpendicular temperature decrease (\(T_{e \perp}/T_{e, 0}<1\)) is observed along the separatrix in the lower left and upper right quadrants.
Additionally, we observe that the regions of perpendicular temperature increase and decrease narrow as the guide field strength increases. 

Recent experimental studies found that e-rec with strong guide fields is accompanied by parallel heating along one separatrix \citep{shi2023using}, which is attributed to the parallel electric field. 
Here we show the electron parallel temperature ratio \(T_{e, \parallel}/T_{e, 0}\) from our three simulations in Fig.~\ref{AT} (h) to (j).
Our results reveal parallel temperature increases along the separatrix in the lower left and upper right quadrants, consistent with experiments \citep{shi2023using}. However, we also observe parallel temperature decreases along the separatrix in the lower right and upper left quadrants.
Furthermore, the parallel temperature increase and decrease are significantly influenced by the magnitude of the guide field. With increasing guide field strength, the parallel temperature increase/decrease regions become thinner and more localized, and the magnitude of the parallel temperature increase/decrease changes notably.
It is also evident from the colorbar in the last two rows of Fig.\ref{AT} that the variations in the parallel temperature ratio \(T_{e, \parallel}/T_{e, 0}\) exceed those in the perpendicular temperature ratio \(T_{e, \perp}/T_{e, 0}\), particularly in case \textbf{C} where the guide fields are stronger.
As a consequence, the parallel temperature ratio $T_{e,\parallel}/T_{e,0}$ shows more correlation with the spatial distribution of $A_e$ than perpendicular temperature ratio \(T_{e,\perp}/T_{e,0}\). Both the spatial distribution of $T_{e,\parallel}/T_{e,0}$ and $A_e$ illustrate a transition from a quadrupolar to a six-polar to an eight-polar structure as the guide field increases.
Specifically speaking, in simulation \textbf{A}, $A_e$ shows a quadrupolar pattern, with two positive bands located on the separatrix in the upper left and lower right quadrants and the other two negative bands along the other separatrix and within the outflow regions.
In simulation \textbf{B}, \(A_e\) exhibits a six-polar structure. In addition to the quadrupolar structure similar to that observed in panel (a), there is a faint band (where $A_e$ is almost zero near \(x\approx 11, \mathrm{y}\approx 16\) along the downstream of the separatrix, which is absent in simulation \textbf{A}. These two faint bands (there is also a faint band in the left-half domain) segregate the negative bands along the separatrix from those within the outflow regions, culminating in a pronounced six-polar structure in both \(A_e\) and \(T_{e,\parallel}\).
In simulation \textbf{C}, an eight-polar structure appears, which features two positive bands on one separatrix paired with two negative bands along the other separatrix. Additionally, there are two bipolar alternations of positive and negative temperature anisotropy bands inside the outflow regions and along the downstream of the separatrix, which is a unique characteristic in this case.
Essentially, our analysis shows that e-rec produces different electron temperature anisotropy patterns under varied guide fields, with more parallel temperature increase/decrease features developing in magnetic islands as the guide field strength increases.

\subsection{Electron velocity distribution functions}\label{part3.4}

In the previous section, we analyzed the electron temperature anisotropy observed in the three cases, highlighting the notable variations that take place as the guide field strength increases. In this section, we further examine the EVDFs at different spatial locations. First, we analyze the EVDFs at the X-point in the three cases. We also focus on regions exhibiting pronounced anisotropy, where we observe a transition from a quadrupole to a six-pole to an eight-pole structure as the guide field intensity increases. 
Fig.~\ref{AT} illustrates the selected positions where EVDFs are calculated, denoted by colored square boxes whose side is about \(\sim \)0.3$d_e$, each containing approximately 500,000 electron macroparticles.
We only sample macroparticles within the bands in the right half of the domain since the structure of the reconnection region is symmetrical, and we find analogous results in the left half of the domain (not shown here).

\begin{figure}
    \centering
    \includegraphics[scale=0.45]{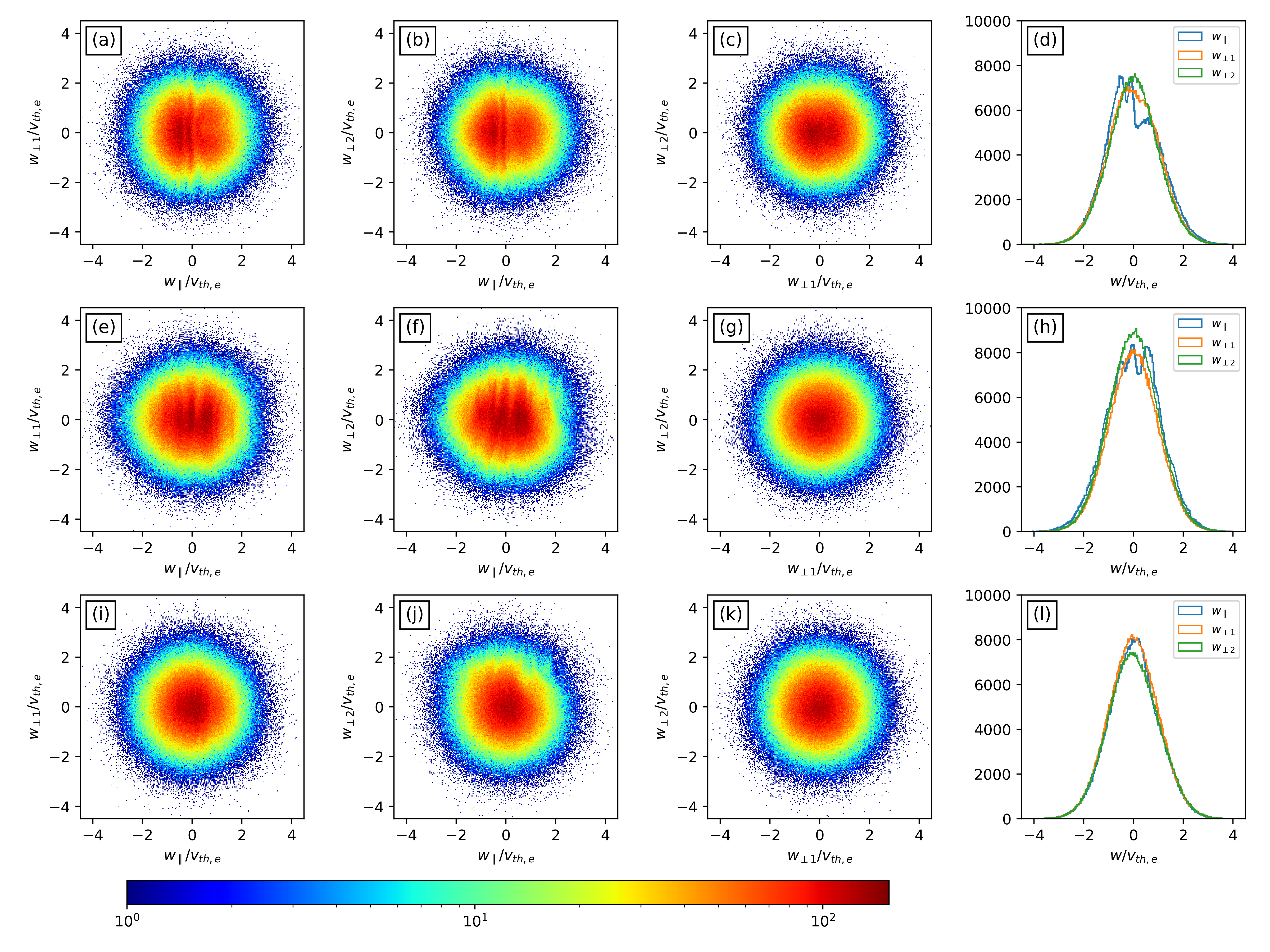}
    \caption{EVDFs in case \textbf{A} are depicted as follows: (a) to (d) correspond to the magenta box (the X-point with negative \(A_e\)); (e) to (h) correspond to the yellow box (the outflow region with negative \(A_e\)); and (i) to (l) correspond to the cyan box (the separatrix in the lower right quadrant with positive \(A_e\)). The colorbar indicates the particle count. The first three columns display the projections of EVDFs in distinct planes, while the last column illustrates the 1D EVDFs, with the y-axis denoting the number of particles.}
    \label{5YVDF}
\end{figure}

Fig.~\ref{5YVDF} presents the 2D projections of the EVDFs at different locations in case \textbf{A}. Each row has four figures, with the initial three depicting the normalized microscopic electron velocity components centered to their mean values, i.e. \(w_{\parallel}/v_{th,e}=(v_{\parallel}-v_{\parallel,mean})/v_{th,e}\), \(w_{\perp1}/v_{th,e}=(v_{\perp1}-v_{\perp1,mean})/v_{th,e}\) and \(w_{\perp2}/v_{th,e}=(v_{\perp2}-v_{\perp2,mean})/v_{th,e}\), respectively. The last panel in each row represents the 1D histogram of these three components with 256 bins.
The velocity components are defined as follows (the same as \citep{goldman2016can,arro2023generation}): $v_{\parallel}$ aligns with the local magnetic field \(\textbf{B}\), $v_{\perp 1}$ points toward \(\textbf{B} \times z\), and $v_{\perp 2}$ is oriented in the direction of \(\textbf{B} \times v_{\perp 1}\). \(v_{\parallel,mean}, v_{\perp1,mean}\) and \(v_{\perp2,mean}\) denote the fluid drift velocity in each direction and have been subtracted. \(v_{th,e}\) represents the initial electron thermal velocity. The colorbar of 2D plots and the y-axis of the 1D histogram indicate the number of macroparticles.

The first row of Fig.~\ref{5YVDF} corresponds to the magenta box (X-point). Panel (a) and (b) indicate evident skewness in \(v_{\parallel}\), implying a net parallel heat flux. This observation is supported by panel (d). We also find the agyrotropy is very small from panel (c) where the projection of the EVDFs on the \(v_{\perp1}-v_{\perp2}\) plane is almost rotationally symmetric. Additionally, in Fig.\ref{AT} (g), parallel heating is evident in this region, corroborated by the slightly irregular oval-shaped EVDFs in panels (a) and (b).
The second row depicts the EVDFs of the yellow box (outflow region). Panels (e) and (f) also display irregular oval-shaped EVDFs, consistent with the parallel heating observed in the outflow region in Fig.~\ref{AT} (g). Skewness is also apparent in panel (h), while panel (g) illustrates small agyrotropy in this region.
Similar gyrotropic characteristics are observed in the cyan box (the separatrix in the lower right quadrant), as depicted in panel (k) of the third row. This region contrasts with the former two, exhibiting perpendicular heating and parallel cooling, as evidenced in Fig.~\ref{AT} (d) and (g). As a consequence, the EVDFs in panels (i) and (j) exhibit slightly irregular vertical oval shapes. Panel (l) indicates that the perpendicular heating is relatively small in this region since the broadening of the perpendicular velocities is not very pronounced with respect to the parallel component.

\begin{figure}
    \centering
    \includegraphics[scale=0.45]{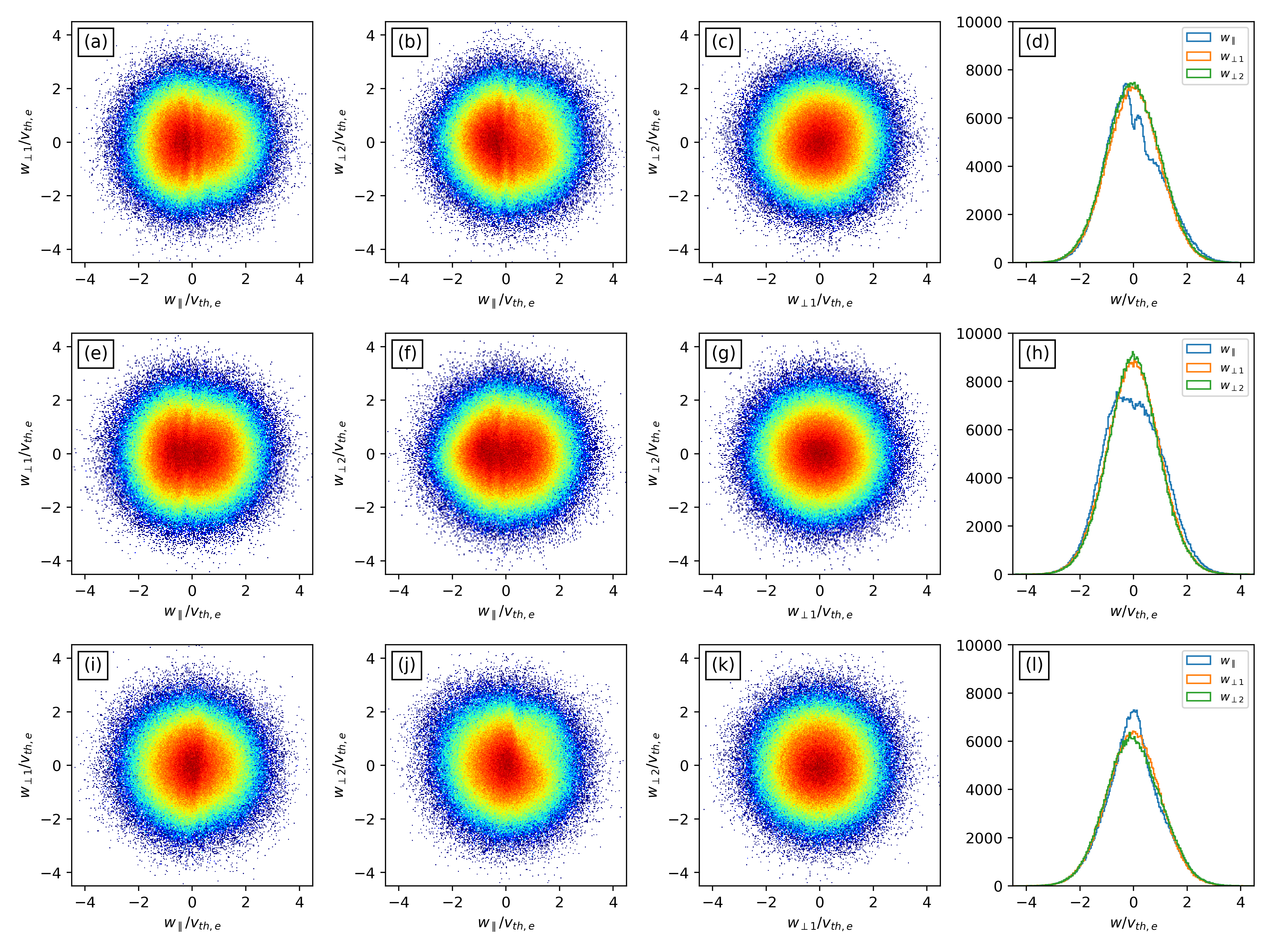}\\
    \vspace{-7.5pt}
    \includegraphics[scale=0.45]{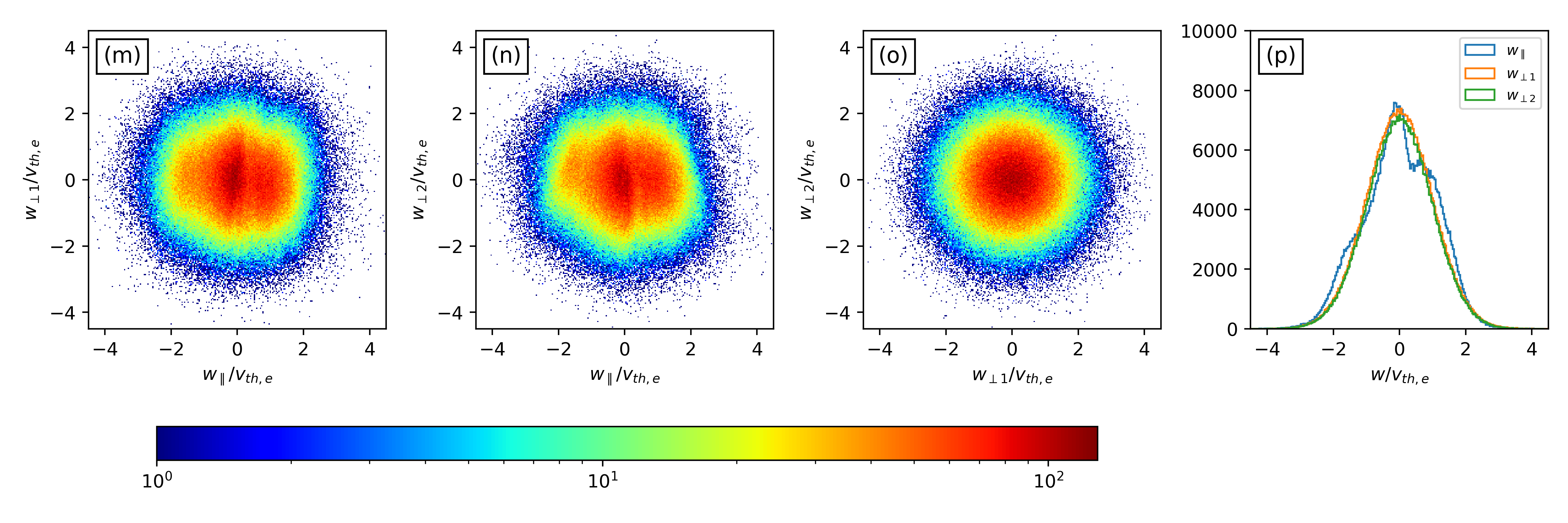}
    \caption{EVDFs in case \textbf{B} are depicted as follows: (a) to (d) correspond to the magenta box (the X-point with negative \(A_e\)); (e) to (h) correspond to the yellow box (the separatrix in the upper right quadrant with negative \(A_e\)); (i) to (l) correspond to the cyan box (the separatrix in the lower right quadrant with positive \(A_e\)); and (m) to (p) correspond to the lime box (the outflow region with negative \(A_e\)). The colorbar indicates the particle count. The first three columns display the projections of EVDFs in different planes, while the last column represents the 1D EVDFs, with the y-axis denoting the number of particles.}
    \label{10YVDF}
\end{figure}

Fig.~\ref{10YVDF} illustrates the 2D projections of the EVDFs at four locations in case \textbf{B}. The first row corresponds to the magenta box (X-point), where the skewness similar to case \textbf{A} is observed in panels (a) and (b). The skewness is also evident in the 1D histogram (panel (d)). Panel (c) displays a nearly gyrotropic distribution, which is also observed for EVDFs calculated inside the other three boxes (panels (g), (k), and (o)) in case \textbf{B}.
The second row shows the EVDFs of the yellow box. Due to the parallel heating in this area, more particles exhibit large parallel thermal velocities. Consequently, panels (e) and (f) display oval-shaped distributions, which are more pronounced than those observed in the yellow box of case \textbf{A}. The parallel heating is also highlighted in panel (h).
On the other hand, the cyan box experiences parallel cooling and perpendicular heating. Therefore, the EVDFs appear elongated in the directions perpendicular to the local magnetic field, as demonstrated in panels (i) and (j). Panel (l) displays the perpendicular heating more distinctly as the two perpendicular components become wider.
The last row demonstrates the EVDFs of the lime box, which also presents prominent parallel heating. Correspondingly, panels (m) and (n) show that 2D EVDFs are elongated in the parallel direction. However, their shapes are irregular, as also shown in panel (p), which is significantly different from the yellow box in case \textbf{A} (also located in the outflow region).

\begin{figure}
    \centering
    \includegraphics[scale=0.45]{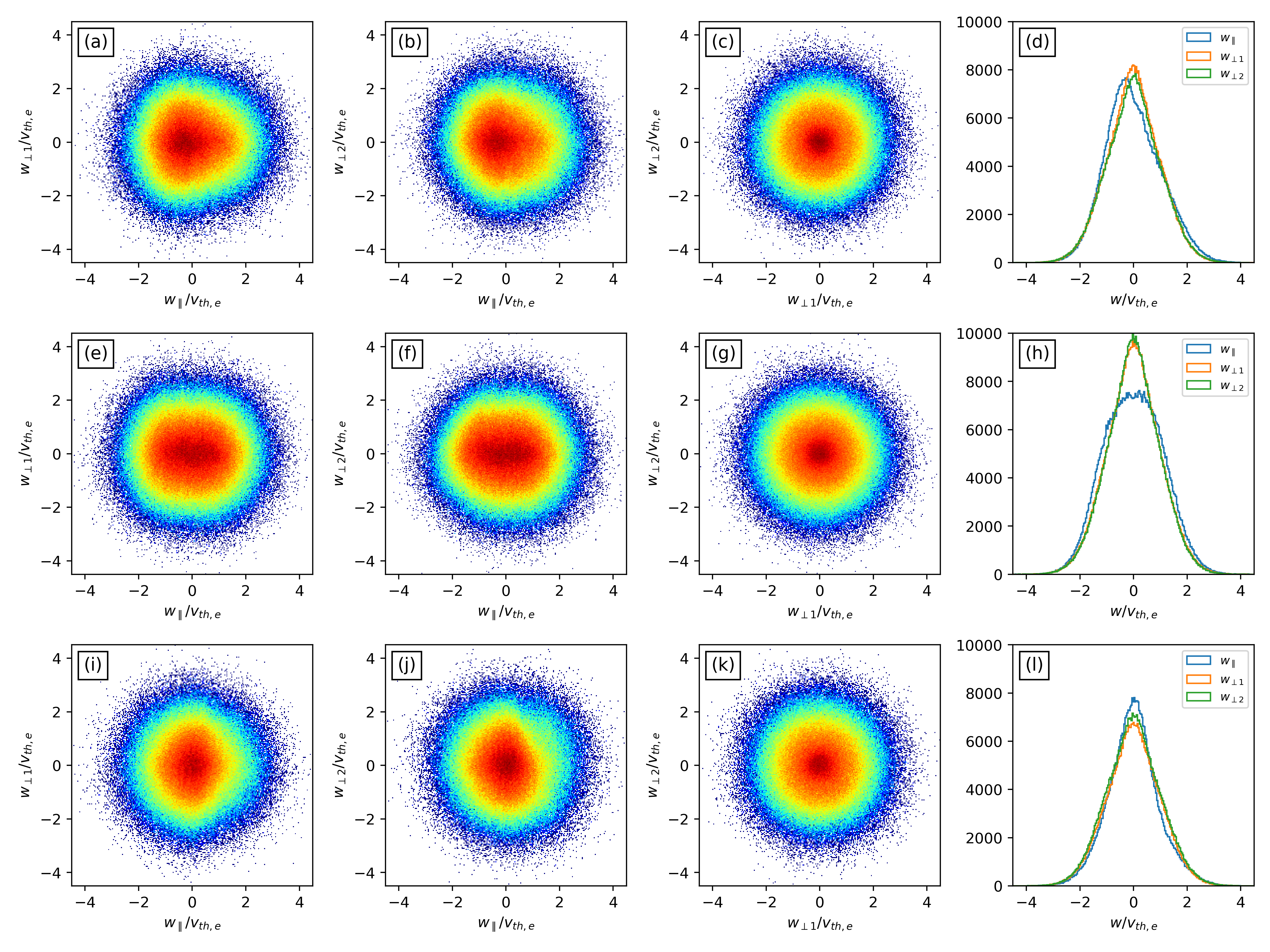}\\
    \vspace{-7.5pt}
    \includegraphics[scale=0.45]{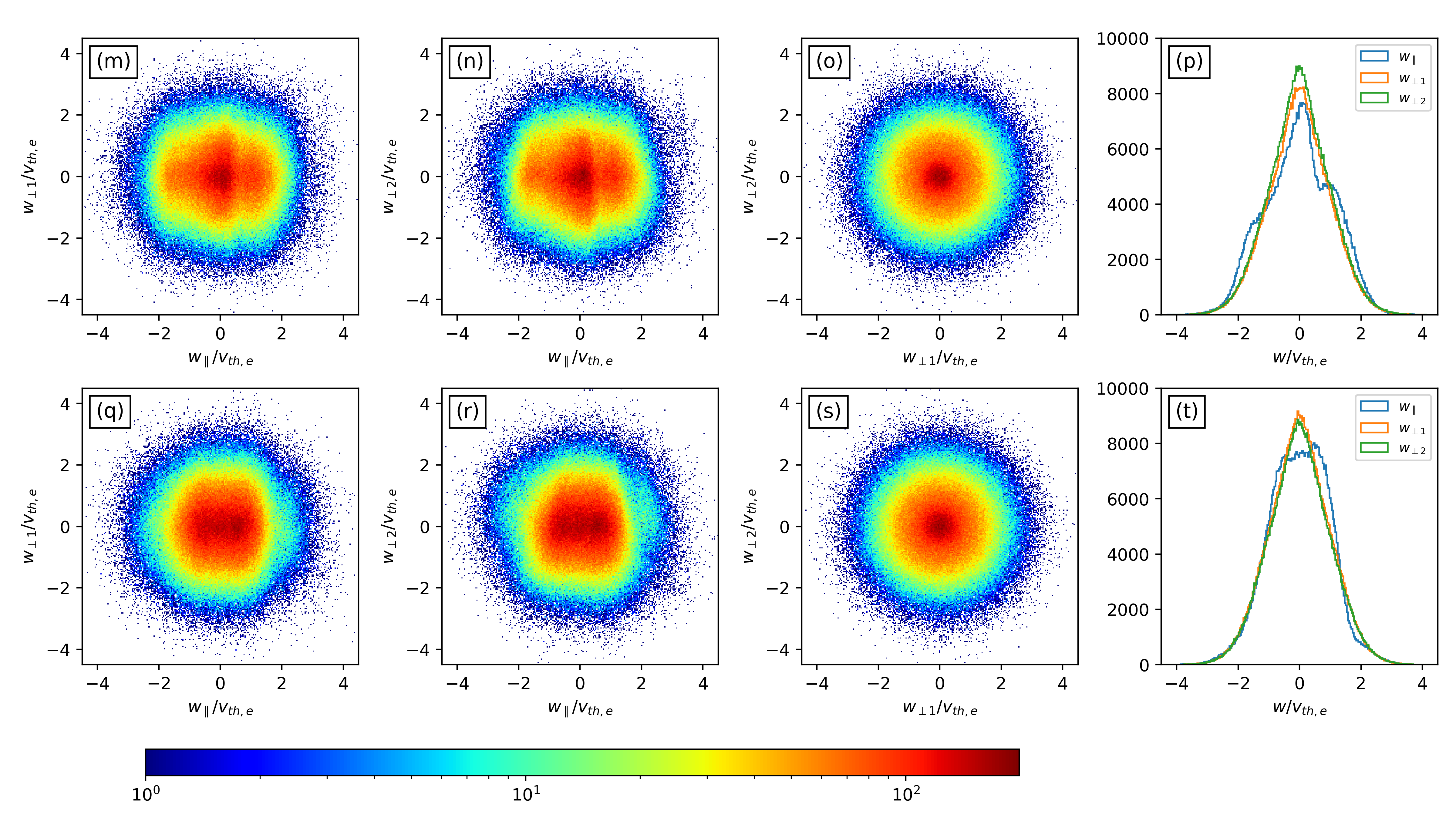}
    \caption{EVDFs in case \textbf{C}. (a) to (d) correspond to the magenta box (the X-point with negative \(A_e\)); (e) to (h) correspond to the yellow box (the separatrix in the upper right quadrant with negative \(A_e\)); (i) to (l) correspond to the cyan box (the separatrix in the lower right quadrant with positive \(A_e\)); (m) to (p) correspond to the lime box (the outflow region with negative \(A_e\)); (q) to (t) correspond to the purple box (the outflow region with positive \(A_e\)). The colorbar indicates the particle count. The first three columns are the projections of EVDFs in different planes. The last column represents the 1D EVDFs, with the y-axis denoting the number of particles.}
    \label{20YVDF}
\end{figure}

In Figure \ref{20YVDF}, we show the 2D projections of the EVDFs at five locations in case \textbf{C}.
The first row illustrates the EVDFs of the magenta box (X-point). Panels (a) and (b) suggest a skewed parallel velocity component, similar to the magenta boxes in cases \textbf{A} and \textbf{B}. Panel (d) represents the 1D EVDFs that confirm this skewness, indicating a significant heat flux near the X-point in case \textbf{C}.
Panel (c) reveals that the EVDFs in the magenta box are nearly gyrotropic, a feature that becomes more notable as the guide field increases. In fact, the EVDFs of all five boxes show gyrotropic features in case \textbf{C}, as shown in panels (g), (k), (o) and (s).
Remarkable anisotropy and parallel heating along the separatrix in the upper right quadrant (the yellow box) is observed in Fig.~\ref{AT} (c) and (i). 
The second row shows the EVDFs of the yellow box. Due to the parallel heating, oval-shaped non-Maxwellian EVDFs appear, as shown in panels (e) and (f). Panel (h) also clearly displays that the parallel velocity component becomes wider.
It is also noted that the oval-shaped EVDFs become more regular and elongated in case \textbf{C}, compared to cases \textbf{A} and \textbf{B}. Correspondingly, the \(T_{e, \parallel}\) of the yellow box in case \textbf{C} also becomes larger than in the other two cases, as shown in Fig.~\ref{AT} (g) to (i).
The third row depicts the EVDFs of the cyan box. Both parallel cooling and perpendicular heating exist along this separatrix in the lower right quadrant, leading to the formation of vertically elongated EVDFs, as illustrated in panels (i) and (j). The parallel cooling is illustrated more clearly in the 1D histograms as \(w_{\parallel}/v_{th,e}\) becomes narrower (panel (l)).
Considering the EVDFs are still gyrotropic in the cyan box, we conclude that the perpendicular heating is isotropic within the plane perpendicular to the local magnetic field. 
The fourth row illustrates the EVDFs of the lime box. Panels (m) and (n) show EVDFs that are stretched along the parallel direction, corresponding to the parallel heating observed in this region in Fig.~\ref{AT} (i). 
In addition, the skewness in panel (p) denotes relatively large heat flux in this area. However, the parallel thermal velocity component becomes very irregular.
The last row demonstrates the EVDFs of the purple box, which exhibit different features from the other boxes. A flat-top parallel velocity distribution is observed for particles with small parallel thermal velocity, while parallel cooling emerges for particles with large parallel thermal velocity, as shown in panels (q), (r) and (t), where 2D EVDFs appear stretched in perpendicular directions and compressed in the parallel direction.

In summary, by investigating the EVDFs in our three runs, we find that the guide field strength has notable effects on the electron parallel velocity component, especially on particles in the outflow regions. 
Different types of EVDFs are observed as the guide field strength increases, and the most various features are obtained in the case of the strongest guide field studied in this work.
Notably, the EVDFs of all the examined locations in the three cases are roughly gyrotropic.

\begin{figure}
    \centering
    \includegraphics[scale=0.395]{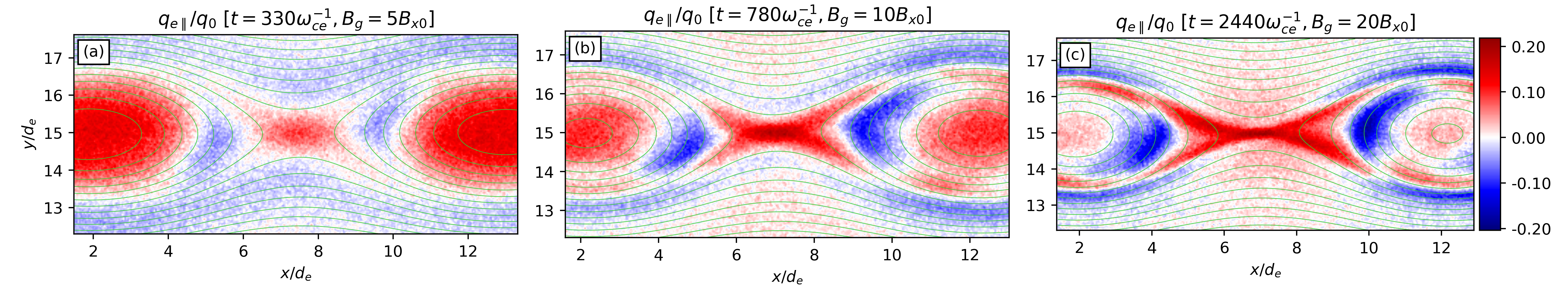}
    \caption{Electron heat flux in the parallel direction (with respect to the local magnetic field) in all three cases, normalized to the reference electron heat flux \(q_0 =n_{0}\,m_e\,v_{th,e}^{3}\). The three figures share the same colorbar, which is located on the far right of the row.}
    \label{qepara}
\end{figure}

Fig.~\ref{qepara} illustrates the electron heat flux in the parallel direction, normalized to the reference electron heat flux \(q_0=n_{0}\,m_e\,v_{th,e}^{3}\) (where $n_0$ is the initial density), for all three runs. These plots show the presence of a strong positive parallel electron heat flux at the X-point, extending to the separatrices as the guide field increases, which is consistent with the EVDFs observed above. Intense heat flux is also evident in the outflow regions, albeit with a direction opposite to that observed in the aforementioned regions.
As the guide field intensifies, the region exhibiting intense heat flux becomes narrower, and the magnitude of the normalized heat flux along the separatrices and within the outflow regions increases. Positive heat flux is also present at the center of the magnetic islands, but its amplitude decreases with increasing guide field strength.

\section{Discussion and Conclusions}\label{part4}

Electron-only reconnection (e-rec) with various strengths of guide fields is systematically investigated by means of 2D fully kinetic Particle-in-Cell simulations, focusing on the anisotropic heating and the EVDFs.
The guide field delays the onset of e-rec and decreases the reconnection rate, similar to what has been observed in previous studies regarding standard ion-scale reconnection \citep{pritchett2005onset,birn2007reconnection}.
The in-plane electron current leads to out-of-plane magnetic field perturbations, consistent with the Hall quadrupolar configuration. Asymmetric quadrupolar electron density structures are also observed, with density enhancements on one separatrix and density cavities along the other. Ions move to form similar density asymmetry structures to neutralize electrons. The motion patterns of electrons and ions are found to be very different. It is noted that the ion outflow jets are absent in our simulations, which is firm proof that the reconnection is in the electron-only regime. 
The disparate velocities and behaviors of electrons and ions are distinctly influenced by the magnitude of guide fields.
Stronger guide fields narrow the density enhancements/cavities and the out-of-plane magnetic field quadrupolar structures. 
This narrowing effect might be related to non-ideal kinetic effects which take place at scales of the order of the electron gyroradius, producing electron-scale fluctuations and structures where electrons eventually decouple from the magnetic field dynamics.
As the guide field intensifies, it is reasonable to think that the non-ideal regions will become thinner since their size is of the order of the electron gyroradius and stronger guide fields imply smaller gyroradii. We plan to investigate this problem in future works.

Temperature anisotropy is observed along separatrices and inside magnetic islands. 
The analysis of \(T_{e, \perp}\) demonstrates that perpendicular temperature increase resides along the separatrix in the upper left and lower right quadrants, and perpendicular temperature decrease is present on the other separatrix. 
Parallel temperature increase, on the other hand, appears on the separatrix displaying perpendicular temperature decrease. It is also noted that parallel temperature decrease manifests on the separatrix showing perpendicular temperature increase. However, the parallel temperature increase/decrease is stronger than the perpendicular temperature increase/decrease in our simulations. The guide field intensity has notable effects on the spatial distribution of the parallel electron temperature.
As the guide field increases, both the regions of parallel and perpendicular temperature increase/decrease become narrower, and parallel temperature increase/decrease exhibits different structures inside magnetic islands, which generates different non-Maxwellian EVDFs in outflow regions. Furthermore, we observe a transition from a quadrupolar to a six-polar to an eight-polar structure in the parallel electron temperature and the temperature anisotropy as the guide field intensity increases. 

EVDFs across different locations are examined in all three cases. The perpendicular velocity components of these three cases exhibit a roughly gyrotropic distribution.
EVDFs near the X-point show velocity distributions with skewness in the direction of the local magnetic field, indicating the existence of relatively large heat fluxes. 
EVDFs with a parallel elongation in the direction of the magnetic field are observed at the separatrices which exhibit parallel heating and perpendicular cooling. EVDFs stretched in the plane perpendicular to the magnetic field appear on the separatrices which show perpendicular heating and parallel cooling.
The guide field is found to have notable effects on the parallel velocity component, especially on electrons in the outflow regions, which is consistent with the observed parallel temperature variations. 
The shapes of the EVDFs are sensitive to the guide field strength and various features develop as the guide field intensifies.
Intense electron heat fluxes in the parallel direction are observed at the X-point, along the separatrices and inside magnetic islands.

The parallel temperature increase observed along one separatrix in our simulations is consistent with a recent experimental study \citep{shi2023using}. However, a discrepancy arises regarding perpendicular temperature increase: while it only emerges along the separatrix in the upper left and lower right quadrants in our simulations, perpendicular heating is observed on both separatrices in experiments. This inconsistency may stem from three-dimensional and/or collisional effects and we plan to pursue further investigations in future studies.

This project has received funding from the ERC Advanced Grant TerraVirtuale (grant agreement No. 1101095310), from the EU Horizon Europe project ASAP (grant agreement No. 101082633), from the KU Leuven Bijzonder Onderzoeksfonds (BOF) under the C1 project TRACESpace, and from the FWO project Helioskill. Views and opinions expressed are however those of the authors only and do not necessarily reflect those of the European Union or the European Commission. Neither the European Union nor the European Commission can be held responsible for them. J. R. is partially supported by the CSC(China Scholarship Council)-IMEC-KU Leuven Scholarship Programme.
We gratefully acknowledge the Gauss Centre for Supercomputing e.V. (www.gauss-centre.eu) for providing computing time on the GCS Supercomputer SuperMUC-NG at Leibniz Supercomputing Centre (www.lrz.de) through the project “Heat flux regulation by collisionless processes in heliospheric plasmas—ARIEL" and "Investigation of suprathermal features in the velocity distribution functions of space and astrophysical plasmas-SupraSpace". M.E.I. acknowledges support from the Deutsche Forschungsgemeinschaft (German Science Foundation, DFG) within the Collaborative Research Center SFB1491.

\bibliographystyle{jpp}

\bibliography{jpp-instructions}

\end{document}